**An image-based transfer learning approach for using i*n situ* processing data to predict laser powder bed fusion additively manufactured Ti-6Al-4V mechanical properties**

Qixiang Luo[1], John D. Shimanek[1], Timothy W. Simpson[2,3], and Allison M. Beese[1,3,*]

[1] Department of Materials Science and Engineering, Pennsylvania State University, University Park, PA 16802

[2] Department of Industrial and Manufacturing Engineering, Pennsylvania State University, University Park, PA 16802

[3] Department of Mechanical Engineering, Pennsylvania State University, University Park, PA 16802

[*] Corresponding author, email: amb961@psu.edu

**Abstract**

The mitigation of material defects from additive manufacturing (AM) processes is critical to reliability in their fabricated parts and is enabled by modeling the complex relations between available build monitoring signals and final mechanical performance. To this end, the present study investigates a machine learning approach for predicting mechanical properties for Ti-6Al-4V fabricated through laser powder bed fusion (PBF-LB) AM using *in situ* photodiode processing signals. Samples were fabricated under different processing parameters, varying laser powers and scan speeds for the purpose of probing a wide range of microstructure and property variations. Photodiode data were collected during fabrication, later to be arranged in image format and extracted to information-dense vectors by the transferal of deep convolutional neural network (DCNN) structures and weights pre-trained on a large computer vision benchmark image database. The extracted features were then used to train and test a newly designed regression model for mechanical properties. Average cross-validation accuracies were found to be 98.7% ($r^2$ value of 0.89) for the prediction of ultimate tensile strength, which ranged from 900 to 1150 MPa in the samples studied, and 93.1% ($r^2$ value of 0.96) for the prediction of elongation to fracture, which ranged from 0 to 17%. Thus, with high accuracy and hardware accelerated inference speeds, we demonstrate that a transfer learning framework can be used to predict strength and ductility of metal AM components based on processing signals in PBF-LB, illustrating a potential route toward real-time closed-loop control and process optimization of PBF-LB in industrial applications.

**Keywords:** Machine learning; Computer vision; Deep convolution neural network (DCNN); Laser powder bed fusion (PBF-LB); *in situ* processing monitoring.

# 1 Introduction

Laser powder bed fusion (PBF-LB) additive manufacturing (AM) is a commonly used layer-by-layer fabrication process for producing metallic parts with complex structures where a laser beam selectively melts 2-dimensional patterns into a layer of a powder bed, with the process repeated through subsequent layers until 3-dimenional components are constructed [1,2]. During fabrication, various types of processing defects and anomalies exist due to the variation of processing conditions, including lack-of-fusion, where incomplete melting of metallic powders form disconnected voids [3,4], keyholing, where an overheated meltpool traps gaseous pores [3], and beading-up, where an over-elongated meltpool breaks up into balling bumps [5,6]. These processing defects severely degrade the mechanical properties of the as-built AM part [7,8]; consequently, detecting and mitigating these defects is an important research area for the advancement of PBF-LB applications and the use of AM in industry.

Post-process inspection techniques, such as X-ray computed tomography [3,9] and other non-destructive testing methods [10,11], have been widely used for the detection of defects. However, the post-fabrication nature and experimental constraints of these approaches preclude their integration into the build system to detect and mitigate defects in real time. In-process or *in situ* monitoring systems provide a path for real-time detection of anomalies or pores [12–15], and therefore a pathway for real-time build analysis, closed-loop feedback control, and process optimization.

*In situ* processing signals representing the thermal condition of the PBF-LB meltpool can serve as an important processing indicator. Such signals have been reported to be correlated with the existence of defect pores and voids [12–14] as well as resultant mechanical properties for as-built samples [15]. For instance, Coeck et al. and Bisht et al. [14,15] studied Ti-6Al-4V by PBF-LB (ProX DMP 320, 3D Systems, Inc., United States) and correlated *in situ* signals captured by Ge photodiode sensors with lack-of-fusion porosity [14] and mechanical properties, including elongation to fracture (EF) and ultimate tensile strength (UTS) [15]. However, no quantitative models were provided to directly relate the *in situ* processing signals to the microstructure or properties.

To investigate the process-structure-property (PSP) relationship of materials, machine learning (ML) related data-driven methods have been widely used in material science and engineering and AM [16,17]. Deep learning (DL) in particular has been used to evaluate image-based *in situ* processing data in AM to detect defects and anomalies using, for example, deep convolutional neural networks (DCNNs) [13,18–20]. In addition, the transfer learning method, facilitated by a growing number of open source databases and GPU-enabled parallel computing, allows for models from the computer vision (CV) field, pre-trained on large-scale datasets such as ImageNet [21] or COCO [22], to be applied to other fields [23–27], reducing computational costs for model retraining and optimization. Multiple recent studies have reported the use of transfer learning of models trained on the ImageNet dataset for the purpose of linking processing signal images to microstructural defects, as a classification task, for PBF-LB AM fabricated materials (316L stainless steel [26], and Inconel 718 [28]).

Despite such advances, direct relationships between processing data and final mechanical properties remain unestablished, especially across multiple processing parameter regimes. Before fabrication systems can incorporate microstructural considerations into feedback-based control, quantitative predictive relationships between *in situ* data and final mechanical properties must be established and be available for calculation within fractions of a second. Models suitable to this goal span from processing to properties and should therefore leverage all information possible to implicitly account for the microstructural basis of this linkage.

Accordingly, in the present study, a DCNN-based transfer learning workflow was used to identify the relationships between processing signals and properties of Ti-6Al-4V fabricated by PBF-LB using a wide range of processing parameters. The *in situ* processing signals were visualized as 2D bitmaps, which were fed into a transfer learning workflow by extracting their features using pre-trained DCNN architectures. The extracted feature vectors were then used to train and test a regression model to predict the mechanical properties of each sample. The primary contributions of the present work include a dataset of layer-wise *in situ* photodiode signal images covering a wide range of processing parameter regimes, the successful transferal of a pre-trained DCNN for image feature extraction, and a fast and accurate mechanical property regression model that illustrates a promising route for real-time closed-loop control and optimization of PBF-LB.

## 2 Materials and Methods

### 2.1 Sample fabrication

To study the PSP relationships of Ti-6Al-4V made by PBF-LB, a total of 42 processing sets were designed using different combinations of processing parameters **(Fig. 1a, Tab. 1)**, with constant hatch spacing (82 μm) and layer thickness (60 μm) but with variation in the laser power P (75-500 W) and scan speed v (400-1480 mm/s). ASTM E8/E8M cylindrical tensile samples were directly fabricated **(Fig. 2)**, using PBF-LB (ProX DMP 320, 3D Systems, Inc., United States), with the axis of each sample aligned with the vertical build direction. After fabrication, samples were stress relieved, but no additional heat-treatments were applied prior to testing. Three samples were fabricated with each parameter set for a total of 126 samples, 120 of which were tested under uniaxial tensile test due to premature failure of 6 samples **(Fig. 1b, 1c)**. Further processing and mechanical testing details can be found in the Supplemental Information and additional characterization in refs [7,29].

Photodiode *in situ* process monitoring was performed using the 3D System Direct Metal Printing monitoring system DMP-meltpool [30] on the ProX DMP 320 PBF-LB machine (**Fig. 3**). To mitigate the angular detection biases for the sensors, the intensity values from both 50 kHz Ge photodiodes were averaged at each recorded position before being rearranged into 2-dimensional (2D) image datasets **(Fig. 4)**. Two datasets were constructed for analysis: (1) an image dataset for which all interior and contour datapoints were retained **(Fig. 4b)**, and (2) one for which the outer contour was removed **(Fig. 4c)** such that only interior points were considered. The Supplemental Information includes more details on the image-based data structure, which captures local intensity variations in photodiode signals that have been found to be qualitatively linked to the processing variations during fabrication [15,30]. Example signal visualizations are shown in terms of samples with different mechanical strength **(Fig. 4e)** and elongation **(Fig. 4f)**.

## 2.2 ML approaches

An image-based regression model was introduced to predict the mechanical property (UTS and elongation to fracture) values for each fabricated sample using the processing signal image datasets. Compared to the heavy computational demands of conventional CNN model application, this image-based workflow was constructed using a novel lightweight DCNN-based transfer learning framework **(Fig. 5)**. The modeling tools were constructed using a deep learning framework with TensorFlow [31] and Keras [32], and graphics processing unit (GPU)-accelerated computation was accomplished using the parallel computing platform by NVIDIA CUDA (Compute Unified Device Architecture) [33]. The Supplemental Information provides details on the computational resources and transfer learning architectures (Table S1).

The multiple convolutional blocks of the pre-tuned DCNN architectures **(Tab. 2)**, allowed for the extraction of representative image feature vectors [16,23]. To study the effect of DCNN architectures on the feature extraction results (i.e., whether image information was fully, partially, or overly extracted), four different architectures were considered: (1) VGG16 [34], (2) ResNet50, (3) ResNet101, and (4) ResNet152 [35]. Each architecture, as provided by the Python Keras library, was individually pre-trained on the ImageNet dataset [32]. Computational complexity increases with the number of CNN layers (shallow to deep: 16, 50, 101, and 152 layers, respectively) and with additional interlayer connections (as exist in the ResNet-structured residual network but not in the VGG-structured plain network).

### 2.2.1 Regression model

The extracted feature vectors were used to construct a regression model to predict the ultimate tensile strength and elongation to fracture of each sample. The architecture of the regressor **(Fig. 5)**, consists of a neural network (NN) structured sequential model where each fully-connected (FC) layer reduces the dimensionality of information until a scalar value is obtained. Through the sequential NN regressor, information from the spatially arranged *in situ* processing data, was smoothly condensed into a single property prediction value. The prediction accuracy for each input 2D layer-wise image was taken to be the mean absolute error (MAE) [32], and regression model parameters were iteratively updated based on this loss function with the Adam optimizer [36]. Details on the model architecture, optimization, and hyperparameters are given in supplementary material **(Tab. S1)**.

### 2.2.2 Performance evaluation

Several error metrics were used to evaluate the performance of mechanical property predictions, including MAE, mean absolute percentage error (MAPE), and accuracy (defined to be 100-MAPE):

$$MAE_i = \frac{1}{M}\sum_{j=1}^{M}|y_{ij} - t_i|, \qquad (Eq.\ 1.)$$

$$\mu_{MAE} = \frac{1}{N}\sum_{i=1}^{N} MAE_i, \qquad (Eq.\ 2.)$$

$$\sigma_{MAE} = \sqrt{\frac{1}{N}\sum_{i=1}^{N}(MAE_i - \mu_{MAE})^2}, \qquad (Eq.\ 3.)$$

$$MAPE = \frac{1}{N}\frac{1}{M}\sum_{i=1}^{N}\sum_{j=1}^{M}\left|\frac{y_{ij} - t_i}{t_i}\right| \times 100\%, \qquad (Eq.\ 4.)$$

where $i$ and $j$ are the index for the sample and the build layer, $N$ is the number of samples, $M$ is the number of build layers within each sample, $t_i$ is the experimental measured mechanical property value for sample $i$, $y_{ij}$ is the predicted property value for build layer $j$ within sample $i$, and $\mu_{MAE}$ and $\sigma_{MAE}$ are the mean and standard deviation values for the absolute prediction error over the entire dataset.

In addition to error metrics that evaluate sample-wise predictions, complementary statistical analyses of correlation and significance were introduced to provide a broader measure of performance over the entire dataset of many samples. For correlation analysis, the $r^2$ value was used to describe the strength of the linear relationship, defined as the square of Pearson's correlation coefficient r [37], between predicted and observed values:

$$r^2 = \left[\frac{\sum_{i=1}^{N}(t_i - \bar{t})(y_i - \bar{y})}{\sqrt{\sum_{i=1}^{N}(t_i - \bar{t})^2}\sqrt{\sum_{i=1}^{N}(y_i - \bar{y})^2}}\right]^2, \qquad (Eq.\ 5.)$$

where $t = \{t_1, t_2, \ldots, t_N\}$ and $y = \{y_1, y_2, \ldots, y_N\}$ are respectively the ground truth and prediction variable sets for a mechanical property, $N$ is the number of samples, and $\bar{t}$ and $\bar{y}$ are the mean values for these sets averaging over all samples within the entire dataset. The $r^2$ value ranges from 0 to 1, where increasing values indicate better predictions. For the significance analysis, the two-sided Student's t-test [37] was applied to determine the statistical significance of the correlation between the ground truth and the predictions of each model. To evaluate the resulting p-values, a significance level of $\alpha = 0.05$ (corresponding to a 95% confidence interval) was set.

## 3 Results
### 3.1 Signal and processing/property relationships

The associations between signal intensity and both processing parameters and mechanical properties were explored with respect to the average signal intensities for each sample as well as to the variation of the signal intensity among all layers within a sample **(Fig. 6)**. In relation to processing conditions, photodiode intensity showed appreciable correlation with laser power ($r^2$ value of 0.52) **(Fig. 6c)**, while the linear energy density did not show notable correlation with photodiode intensity ($r^2$ value nearly 0) **(Fig. 6b)**. Regarding mechanical properties, both UTS and EF showed very slight correlations with photodiode intensity ($r^2$ value of 0.01 and 0.02 for strength and elongation, respectively) **(Fig. 6e, 6f)**. Such low linear association metrics highlight the limited utility of spatially averaged values in the representation of in-process signals for mechanical performance modeling.

Another important aspect of the signal intensity is the uncertainty, generated by the variation of intensity values along different build layers and shown as the radius of the sample data point **(Fig. 6a, 7d)**, and error bars **(Fig. 6b-f)**. The photodiode intensity had high variability in processing regimes where defects were present, as shown in the processing parameter space **(Fig. 6a)** at the lower left (lack-of-fusion), upper left (keyholing), and upper right (beading-up) regions. This is indicative of large signal fluctuations corresponding to the generation of large and irregular processing pores. While average processing signal intensity is indicative of the processing conditions, especially laser power, the spatial signal variability contains much of the information related to defect formation and is therefore the critical input to the proposed ML approach.

### 3.2  ML model performance

Results of the overall model performance in terms of predicting mechanical properties for different combinations of DCNN architectures and images with and without contours using the transfer learning-based ML approach are shown in **Figure 7** and detailed in the supplementary material **(Fig S1, S2)**. The shape of each learning curve, reported as epoch versus performance in the supplementary material **(Fig. S3)**, indicates the prediction results are optimized, with no signs of underfitting or overfitting. Model performance was evaluated using the average accuracy value over five-fold cross-validation testing. In addition to the proposed transfer learning workflow of combining transferred pre-trained DCNN architectures with a NN regression model trained in the present work, several baseline models with conventional ML architectures were tested as comparisons, including different combinations of feature extractors (transferred, non-transferred) and regressors (linear, nonlinear, NN), with results summarized in **Table 3**, and discussed in subsection 3.2.1.

With its greater complexity, the ResNet architecture outperformed the VGG architecture in prediction of both tested properties. For both VGG and ResNet architecture, the dataset without contours resulted in a notable performance drop compared to the dataset with contours. Among all cases, the ResNet101 architecture and input of images with contours achieved the highest accuracy in predicting mechanical properties, with prediction accuracies of 98.7% and 93.1% ($r^2$ values of 0.87 and 0.96), and absolute errors of 12.98 MPa and 0.61%, for UTS and EF, respectively. These accuracies can be put into the context of experimental errors by considering the distribution of mechanical properties, which can be defined as the property variance among sets of samples fabricated under the same PBF-LB system and processing conditions **(Tab. 1)**. The experimental mean standard deviation of 12.07 MPa in strength and 0.79% in elongation puts the model predictions within the range of experimental measurement variations, supporting model validity.

### 3.2.1 Baseline models

The two types of baseline models correspond to changes in each of the two stages of the proposed transfer learning framework: feature extraction and regression modeling. All the combination of methods along with their performance in predicting mechanical strength and elongation have been summarized **(Tab. 4)** based on the same dataset of signal image data that included the contour region.

To explore the effect of transferring pre-trained DCNN models for use in predicting mechanical properties, a more direct approach was constructed using principal component analysis (PCA) to reduce the dimensionality of input image data from high (150k dimensions due to the flattening of each pixel and RGB channel in each image) to low (chosen to be 500 dimensions) with a minimum loss of information. The output PCA-based (non-transfer learning) and DCNN-based (transfer learning) feature representations were separately used to train a linear regression (LR) model to predict mechanical properties. Within the LR model, the DCNN-extracted features achieved better performance than PCA-extracted features in predictions of both strength (accuracy of 96.9% compared to 96.3%, $r^2$ value of 0.49 compared to 0.01) and elongation (accuracy of 75.5% compared to 20.1%, $r^2$ value of 0.85 compared to 0.52). Furthermore, the t-test of UTS predictions using PCA-based features resulted in a p-value of 0.42, indicating that the model result was not statistically significant. The preceding results suggest that more information was captured through the transferred DCNN-based feature extraction method than through the non-transfer PCA method, leading to better performance even within a simple LR model.

To consider the impact of regression model choice, three commonly used model types were selected as baseline comparisons against the NN model: linear regression (LR) [37], support vector machine (SVM) [38], and random forest (RF) [39]. Each regression model used the feature vector extracted from the ResNet101 architecture as input to predict strength and elongation. Comparing all baseline regression models **(Tab. 4)**, predictive power increased with model complexity, as seen in the NN's outperformance of the three simpler models (LR, SVM, RF) in predictions of both strength (accuracy increases of 1.8%, 0.7% and 0.6%, $r^2$ increases of 0.4, 0.2, and 0.09) and elongation (accuracy increases of 17.6%, 21.7% and 13.2%, $r^2$ increases of 0.11, 0.14, and 0.05). Details of the model configurations and performance are given in the supplementary material.

## 4 Discussion

The influence of model architectures, including both structure and depth, on the predictive power and computational inference cost is discussed in the first of the following sections. What physical information becomes encoded into the model input data also plays a critical role in its performance, and so the role of processing signal data from the contour region is also discussed.

### 4.1 DCNN architectures

To study model performance on predicting mechanical properties over a range of values (i.e., ultimate tensile strength varied from 900-1150 MPa, elongation to fracture varied from 0-17%), comparisons between the measured properties and those predicted by in *in situ* data are made, given in details in supplementary material **(Fig. S1, S2)** for image datasets without and with contours, respectively. The axes correspond to the measured ground truth results (x) and the model predictions (y), while the diagonal dashed line illustrates the ideal case where the prediction equals ground truth. The solid curve provides the predicted UTS or EF for each sample versus the measured property. In these figures, the prediction value is calculated by averaging the predictions over all build layers within the sample. The layer-wise prediction residuals showed no systematic variation along the gauge region of each sample, so their overall variation is shown as the shaded error bands around the sample-wise prediction curve. The closer the solid curve is to the dashed line, the higher the accuracy of the prediction; additionally, the narrower the error band, the lower the uncertainty for each sample. For each subplot, the final prediction performance over the entire probed property range is provided in terms of error metrics (MAE and MAPE) and statistical metrics ($r^2$ value).

The impact of different DCNN architectures is attributed to the different structure designs (VGG and ResNet architecture), and the depth of architectures (16, 50, 101, and 152 layers) **(Tab. 2)**. Compared to the VGG-structured network, which includes multiple convolutional (Conv) layers into contiguous blocks to hierarchically extract image features from shallow to deep, the ResNet-structured residual network created additional shortcuts between non-adjacent CNN layers (layer 1 and 3 within each ResNet Conv block) called identity mappings. These connections restore feature information and pass it to subsequent layers, preventing the feature from being over-extracted. As result, without the addition of skip connections between layers in the convolutional blocks, features may be over-extracted and therefore result in underperforming when used as inputs to the regression model.

Comparing the performance of the studied DCNN architectures **(Fig. S1, S2)**, the ResNet designs achieve better performance than VGG (16 layers) design. The overall prediction performance in ultimate tensile strength was better with ResNet than VGG (accuracy of 98.5% compared to 98.3%, and $r^2$ value of 0.86 compared to 0.81, averaging image datasets with and without contours), and the same trend was also found in the predictions of elongation to failure (accuracy of 91.2% compared to 89.0%, and $r^2$ value of 0.94 compared to 0.92).

Examining each individual prediction value **(Fig. S1, S2)**, VGG and ResNet architecture results show similar overfitting bias toward ranges containing the majority of the data (89% and 73% of data fell in ranges of 950-1150MPa UTS, and 6-15% EF, respectively) with the penalty being underfitting of sparse data regions (11% and 27% of data within ranges of < 950MPa and 1100-1250MPa UTS, and < 6% and 15-17% EF, respectively). These favored overfit ranges are distinguished by shorter distances between the (solid) prediction curve and the (dashed) ideal line, indicating a higher prediction accuracy in sample-wise properties **(Fig. S1, S2)**. Computationally, running the DCNN architectures on a GPU resulted in fast data processing, with an average processing speed of 10-54 ms per image for model testing and 17-58 ms for training, providing a promising avenue for real-time process monitoring and control during PBF-LB AM **(Fig. 8)**.

Comparing the DCNN architecture depths, the model performance on the same ResNet architecture design was tested with different architecture depths (50, 101, and 152 convolution layers). Among the three ResNet depths evaluated, ResNet101 was found to perform best, balancing fitting and generalizability of the model, yielding an average overall accuracy of 98.6% ($r^2$ value of 0.87) in predicting UTS and 91.9% ($r^2$ value of 0.95) in predicting EF for both image datasets (i.e., with and without contours). Increasing model complexity can lead to overfitting on training data with poor performance on the unseen test dataset, while decreasing model complexity can lead to underfitting due to incomplete learning/interpretation of both training and test datasets.

4.2   Processing signal features

Within a DCNN-based transfer learning workflow, the processing signal features **(Fig. 4c)**, played an important role as image characteristics that indicate sample processing conditions and resultant mechanical properties. To quantify the contribution of different features on model performance, two different types of image features were investigated: (1) hatch pattern signals, originating from the interior region of each layer, and (2) hatch-contour pattern signals, originating from the outer contour or near the contour-hatch intersection region. Literature studies [40,41] have reported that hatch-contour regions can give rise to processing pores, generated by repeated heating due to overlapping of parallel hatches and outer contours, diminishing the mechanical properties of the fabricated parts.

Comparing both the overall performances **(Tab. 3)**, and the regression model performance over the range of each predicted property **(Fig S1, S2)**, it was concluded that the removal of contours and near contour features damaged the integrity of input processing signal data. As a result, the overall model performance for the full image dataset, averaged over all ResNet architectures, outperformed the partial image dataset (without contour data) by 0.3% and 2.8% in cross-validation accuracies and by 0.04 and 0.02 in $r^2$ value, for predicting strength and elongation. A similar trend in terms of the layer-wise uncertainty was also found when predicting strength and elongation properties, illustrated by the narrower shaded error bands for the full image dataset **(Fig. S2)** compared to the partial image dataset **(Fig. S1)**. Based on the improved performance and significance metrics, contour and near contour features should be taken into consideration as important processing signals.

The removal of outer contour data resulted in the shrinkage of the sample cross-section area (reduced by 19%), meaning it was possible that the prediction accuracy drop was simply due to the shrinkage of the overall sample size regardless of the utility of the removed information. To isolate this effect, a third image dataset was created by cropping the same amount of area at one side of the sample, covering both bulk and contour regime features. The cropped dataset resulted in similar performance in predicting both strength and elongation properties, relative to the partial image dataset, which confirmed that data reduction was not the key factor leading to an accuracy drop. Instead, the contour features were confirmed to contain important information.

5   Conclusions

In the present work, a DCNN-based transfer learning model was introduced to identify the relationship between *in situ* processing signals collected by photodiodes during fabrication and the resultant ultimate tensile strength and elongation to failure for PBF-LB additively manufactured

Ti-6Al-4V. The main conclusions and contributions are the following:
- Layer-wise photodiode processing signals, covering a wide range of fabrication conditions, were linked with sample-wise mechanical properties to form a dataset suitable to test data-based modeling workflows.
- The effective transfer of a pre-trained DCNN image encoder was demonstrated through improved model performance compared to a non-transfer learning, PCA-based approach, indicating that the feature extraction process successfully captured aspects of the spatial photo-intensity critical to mechanical performance.
- The NN-based regression model, trained on extracted image feature vectors, achieved accuracies of 98.7% ($r^2$ value of 0.89) and 93.1% ($r^2$ value of 0.96) in predicting ultimate tensile strength and elongation to fracture, respectively.
- Supported by GPU hardware acceleration, such a high accuracy, rapid processing, and highly transferable workflow to predict mechanical properties during fabrication via *in situ* process monitoring provides a means for eventual closed-loop process control and optimization in PBF-LB AM.

**Figures**

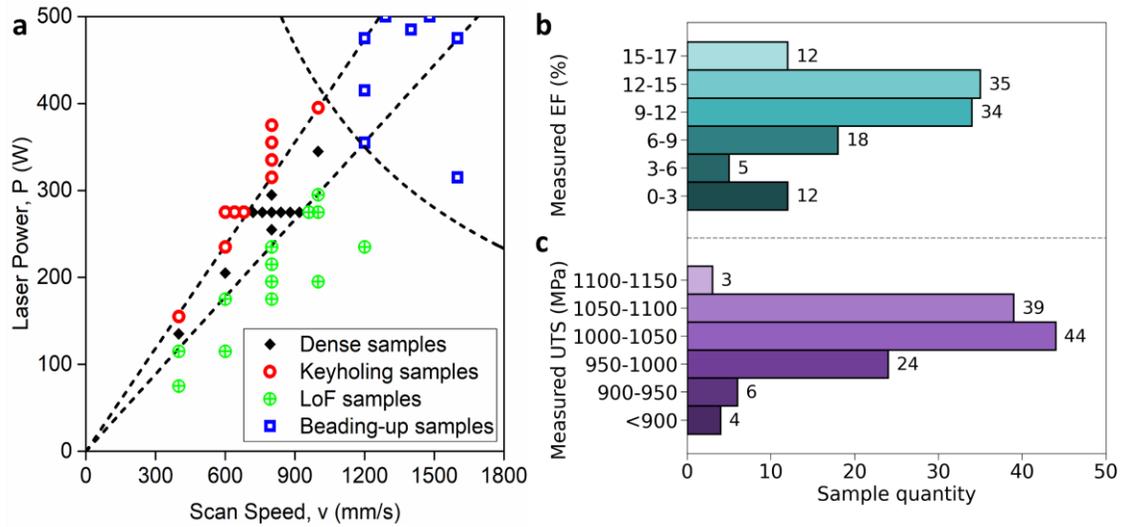

**Figure 1.** (a) Process map with symbols corresponding to the 42 combinations of laser power and scan speed that were investigated to fabricate Ti-6Al-4V by PBF-LB. A total of three samples were made for each unique parameter set for a total of 126 samples. Resultant measured (b) ultimate tensile strength and (c) elongation to fracture distributions for the fabricated samples.

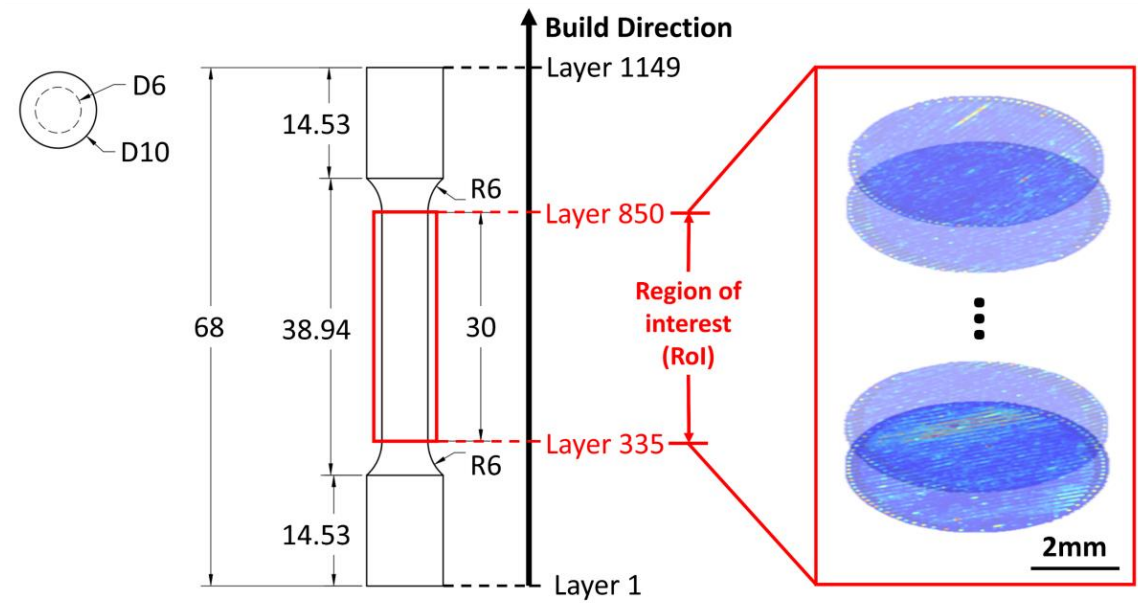

**Figure 2.** ASTM E8/E8M cylindrical sample geometry (all dimensions in mm), with region of interest (ROI) volume of the present study highlighted at the gauge region. *In situ* processing signals were collected for each build layer during fabrication by recording the laser position and corresponding photodiode signal intensity with a frequency of 50 kHz throughout the entire build. The layer-wise data were segmented into data for individual samples, then visualized into intensity colormap images that comprise the photodiode image dataset.

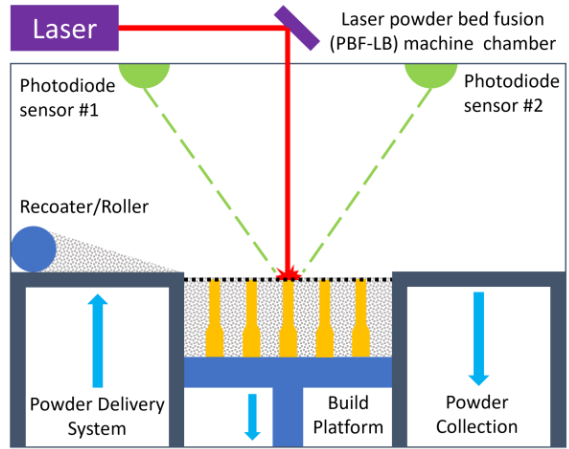

**Figure 3.** Schematic of the PBF-LB system with *in situ* processing monitoring via two photodiode sensors at the top of the chamber.

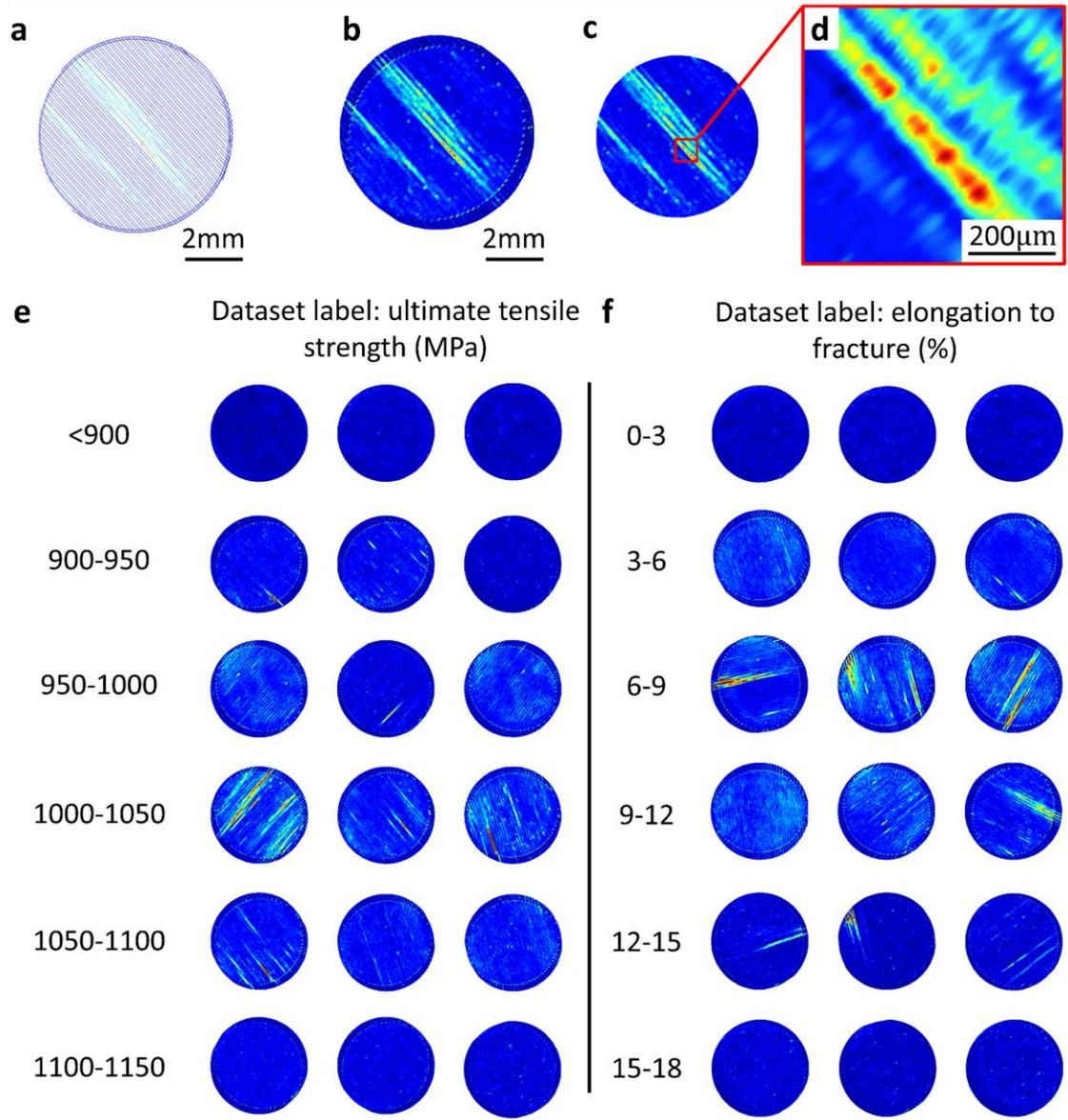

**Figure 4.** Photodiode signal intensity maps for a single layer of a sample illustrating the raw signal data and methods used for visualization: (a) raw photodiode signal data where each point corresponds to the photodiode signal at the time the laser was in that location, (b) 2D bitmap visualization of a layer including interior and outer contour, (c) 2D bitmap visualization of a layer including interior but no outer contour, and (d) zoomed in view of the signal details. Example images of photodiode data representing different ranges of (e) ultimate tensile strength and (f) elongation to fracture.

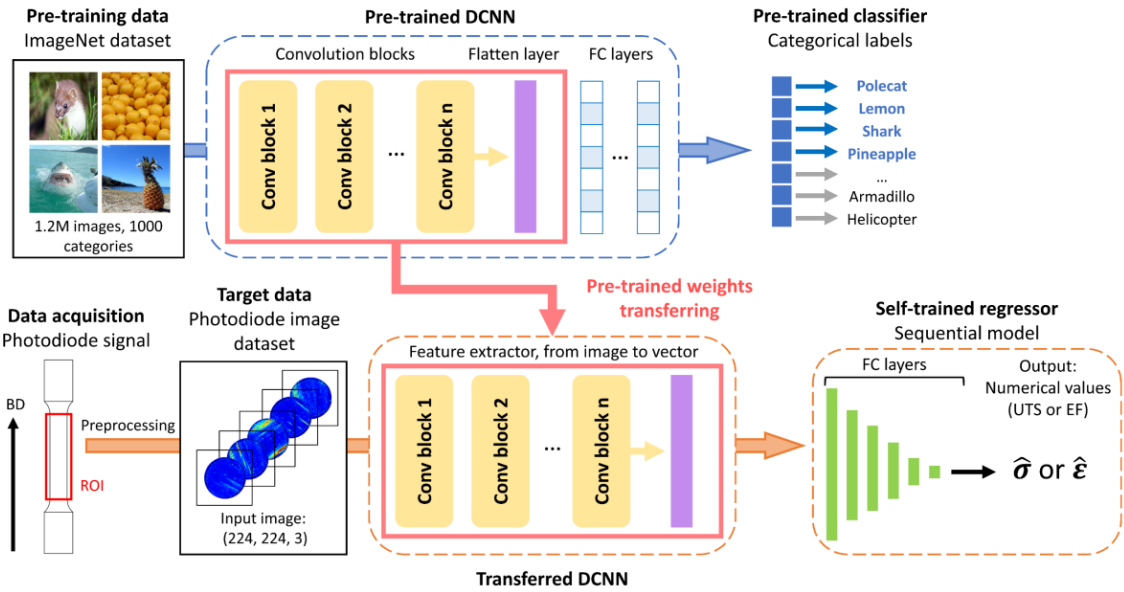

**Figure 5.** Schematic of the DCNN-based transfer learning workflow for predicting mechanical properties of Ti-6Al-4V by PBF-LB. A pre-trained DCNN model and weights were transferred from the ImageNet dataset (sample images randomly selected and presented for demonstration). This DCNN model was used to extract features from each photodiode image collected in the present study into vectorized feature datasets. Collections of these feature vectors for each sample were then used to train a regression model to predict mechanical properties.

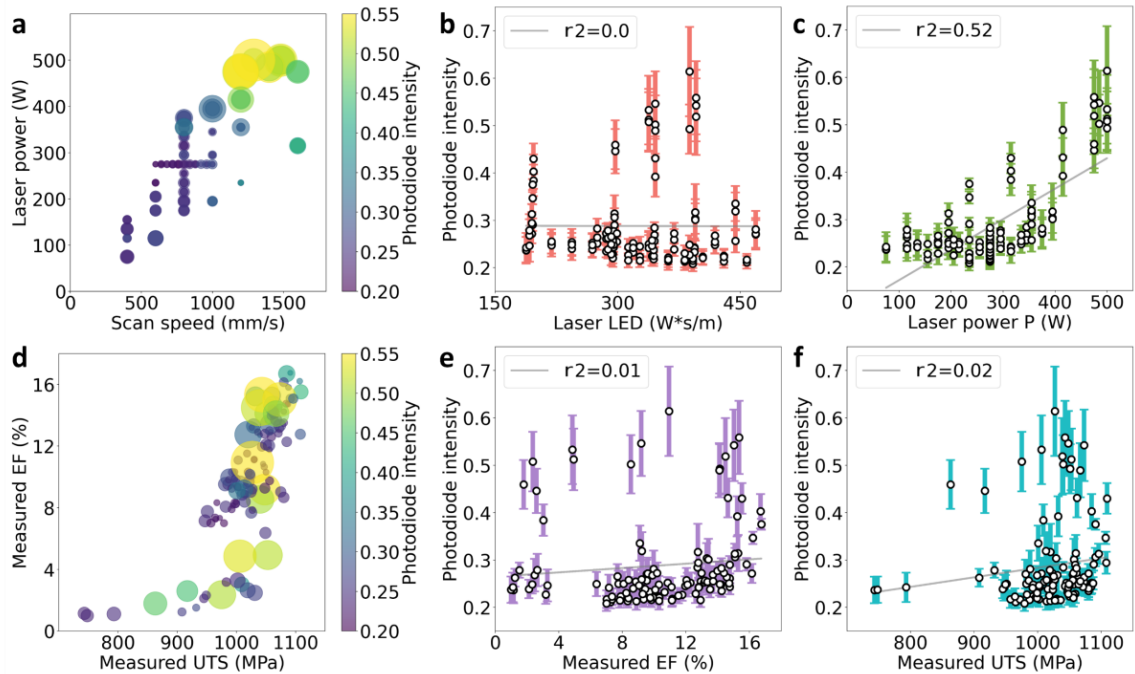

**Figure 6.** Relationships among *in situ* processing signals, processing conditions and mechanical properties. (a, d) Photodiode intensity for each processing condition tested (a) versus processing conditions and (b) versus mechanical properties, where the color and size of each point is the mean and standard deviation of the signal intensity for each sample. (b, c) Correlation between photodiode intensity and processing features (linear energy density and laser power), (e, f) correlation between photodiode intensity and mechanical properties (elongation to fracture and ultimate tensile strength).

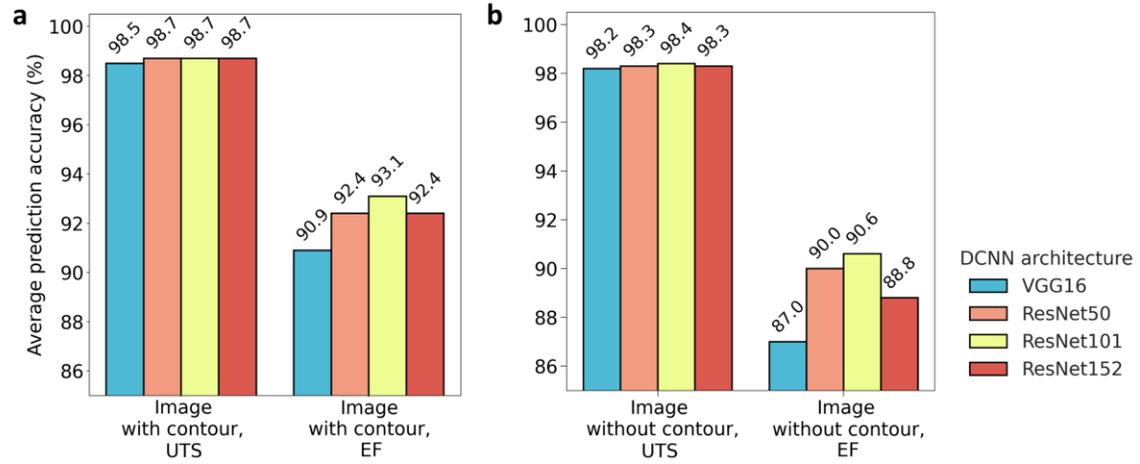

**Figure 7.** Summary of the performance of each model applied in predicting the ultimate tensile strength and elongation to fracture, using different DCNN architectures. Model results for dataset of: (a) images with contours, and (b) image without contours.

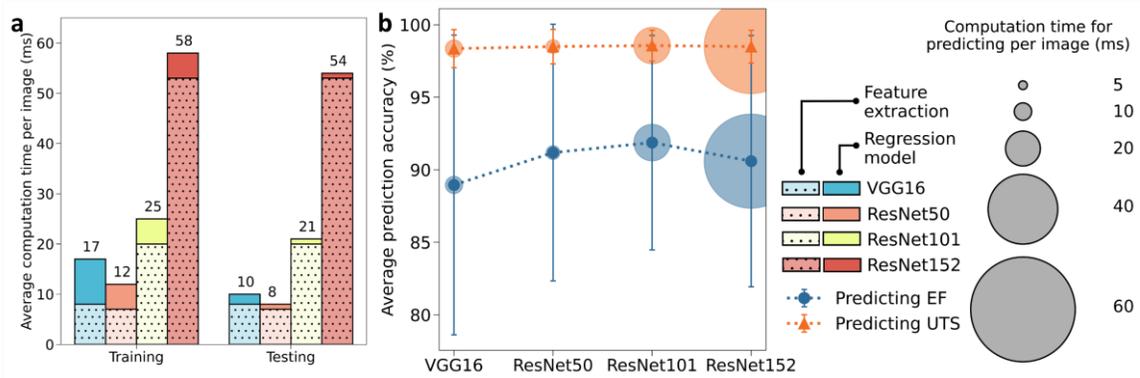

**Figure 8.** Computational cost of the transfer learning modeling. (a) Comparison of the computation time for the training and testing of each input image, using the four explored DCNN architectures (VGG16, ResNet50, ResNet101, and ResNet152), sectioned into the two primary processing steps of feature extraction and regression prediction. (b) Comparison between computational cost (represented by the size of the circle) and the regression model performance (represented by the mean and standard deviation of prediction accuracy over all tested samples) for different DCNN architectures.

# Tables

**Table 1.** Processing parameters and resultant mechanical properties for the fabricated PBF-LB AM Ti-6Al-4V samples.

| Index | Laser Power P (W) | Scan Speed v (mm/s) | Ultimate tensile strength (MPa) | | | Elongation to fracture (%) | | |
|---|---|---|---|---|---|---|---|---|
| | | | set1 | set2 | set3 | set1 | set2 | set3 |
| 1 | 275 | 800 | 1030 | 1018 | 1024 | 11.1 | 11.5 | 10.1 |
| 2 | 275 | 760 | 1020 | 997 | 1002 | 10.3 | 9.9 | 9.3 |
| 3 | 275 | 720 | 1014 | 987 | 999 | 8.9 | 9.8 | 8.2 |
| 4 | 275 | 680 | 992 | 997 | - | 7.9 | 10.3 | - |
| 5 | 275 | 640 | 994 | 967 | 964 | 8.1 | 8.3 | 7.4 |
| 6 | 275 | 600 | 965 | 972 | - | 7.3 | 7.3 | - |
| 7 | 275 | 840 | 1050 | 1030 | 1043 | 10.3 | 11.1 | 12.0 |
| 8 | 275 | 880 | 1056 | 1036 | 1071 | 11.8 | 12.9 | 12.8 |
| 9 | 275 | 920 | 1058 | 1056 | - | 12.9 | 13.0 | - |
| 10 | 275 | 960 | 1064 | 1080 | 1061 | 14.8 | 14.2 | 14.4 |
| 11 | 275 | 1000 | 1087 | 1073 | 1061 | 14.7 | 14.6 | 14.2 |
| 12 | 175 | 800 | 1085 | 1075 | 1067 | 12.3 | 9.0 | 13.7 |
| 13 | 195 | 800 | 1080 | 1073 | 1064 | 8.7 | 13.4 | 14.1 |
| 14 | 215 | 800 | 1080 | 1073 | 1050 | 13.5 | 13.1 | 12.1 |
| 15 | 235 | 800 | 1059 | 1056 | 1055 | 12.6 | 9.8 | 12.0 |
| 16 | 255 | 800 | 1045 | 1047 | 1048 | 10.8 | 10.7 | 12.8 |
| 17 | 295 | 800 | 1016 | 1027 | 1023 | 9.4 | 9.2 | 10.2 |
| 18 | 315 | 800 | 1003 | 992 | 1021 | 11.1 | 9.3 | 9.7 |
| 19 | 335 | 800 | 993 | 1010 | 999 | 10.3 | 8.8 | 9.9 |
| 20 | 355 | 800 | 986 | 1013 | 1001 | 9.5 | 9.2 | 9.1 |
| 21 | 375 | 800 | 981 | 1006 | 988 | 9.5 | 9.6 | 10.0 |
| 22 | 75 | 400 | 743 | 748 | 793 | 1.2 | 1.0 | 1.1 |
| 23 | 115 | 600 | 1002 | 1031 | 1019 | 3.3 | 2.5 | 2.7 |
| 24 | 195 | 1000 | 1080 | 1108 | 1096 | 16.2 | 12.8 | 15.1 |
| 25 | 235 | 1200 | 1088 | 1107 | 1091 | 15.8 | 16.2 | 16.8 |
| 26 | 315 | 1600 | 1085 | 1009 | 1109 | 16.7 | 3.0 | 15.5 |
| 27 | 115 | 400 | 945 | 908 | 932 | 2.1 | 1.2 | 1.5 |
| 28 | 175 | 600 | 1048 | 1049 | 1025 | 9.3 | 6.4 | 7.7 |
| 29 | 295 | 1000 | 1084 | 1057 | 1059 | 14.8 | 14.2 | 14.0 |
| 30 | 355 | 1200 | 1059 | 1060 | 1091 | 14.8 | 15.3 | 15.1 |
| 31 | 475 | 1600 | 917 | 863 | - | 2.6 | 1.8 | - |
| 32 | 135 | 400 | 951 | 982 | 947 | 7.8 | 8.1 | 7.2 |
| 33 | 205 | 600 | 998 | 1024 | 1014 | 8.3 | 9.0 | 8.4 |
| 34 | 345 | 1000 | 1040 | 1055 | 1026 | 12.7 | 13.7 | 13.2 |
| 35 | 415 | 1200 | 1032 | 1062 | 1067 | 15.2 | 14.6 | 14.1 |
| 36 | 485 | 1400 | 1048 | 1040 | - | 9.2 | 8.5 | - |
| 37 | 500 | 1480 | 1053 | 1006 | 975 | 4.9 | 4.9 | 2.4 |
| 38 | 155 | 400 | 957 | 983 | 949 | 7.0 | 3.2 | 2.3 |
| 39 | 235 | 600 | 979 | 1027 | 999 | 7.0 | 7.9 | 8.4 |
| 40 | 395 | 1000 | 1029 | 1057 | 1020 | 13.3 | 13.4 | 12.8 |
| 41 | 475 | 1200 | 1039 | 1073 | 1043 | 14.5 | 15.0 | 15.3 |
| 42 | 500 | 1290 | 1051 | 1027 | - | 14.1 | 10.9 | - |

**Table 2**. Summary of the DCNN architectures used in the present study, including VGG16, ResNet50, ResNet101 and ResNet152. The letters stand for the architecture structure (VGG or ResNet) and the numbers stand for the total number of layers with trainable parameters. Different type and number of layers for each architecture are summarized in terms of convolutional (Conv, with trainable parameters), fully-connected (FC, with trainable parameters), and pooling (max and average-pooling, without trainable parameters).

| DCNN architecture | VGG | DCNN type | Plain network | | | | |
|---|---|---|---|---|---|---|---|
| Block name | Layer name | Kernel shape | Stride shape | Feature shape | Layer quantity | | |
| | | | | | VGG16 | | |
| | Input image | - | - | (224, 224, 3) | - | | |
| Conv b1 | Conv l1-1 | (3, 3, 64) | (1, 1) | (224, 224, 64) | 1 | | |
| | Conv l1-2 | (3, 3, 64) | (1, 1) | (224, 224, 64) | 1 | | |
| | Max-pooling | (2, 2) | (2, 2) | (112, 112, 64) | 1 | | |
| Conv b2 | Conv l1-1 | (3, 3, 128) | (1, 1) | (112, 112, 128) | 1 | | |
| | Conv l1-2 | (3, 3, 128) | (1, 1) | (112, 112, 128) | 1 | | |
| | Max-pooling | (2, 2) | (2, 2) | (56, 56, 128) | 1 | | |
| Conv b3 | Conv l3-1 | (3, 3, 256) | (1, 1) | (56, 56, 256) | 1 | | |
| | Conv l1-1 | (3, 3, 256) | (1, 1) | (56, 56, 256) | 1 | | |
| | Conv l1-2 | (3, 3, 256) | (1, 1) | (56, 56, 256) | 1 | | |
| | Max-pooling | (2, 2) | (2, 2) | (28, 28, 256) | 1 | | |
| Conv b4 | Conv l4-1 | (3, 3, 512) | (1, 1) | (28, 28, 512) | 1 | | |
| | Conv l1-1 | (3, 3, 512) | (1, 1) | (28, 28, 512) | 1 | | |
| | Conv l1-2 | (3, 3, 512) | (1, 1) | (28, 28, 512) | 1 | | |
| | Max-pooling | (2, 2) | (2, 2) | (14, 14, 512) | 1 | | |
| Conv b5 | Conv l5-1 | (3, 3, 512) | (1, 1) | (14, 14, 512) | 1 | | |
| | Conv l1-1 | (3, 3, 512) | (1, 1) | (14, 14, 512) | 1 | | |
| | Conv l1-2 | (3, 3, 512) | (1, 1) | (14, 14, 512) | 1 | | |
| | Max-pooling | (2, 2) | (2, 2) | (7, 7, 512) | 1 | | |
| FC block | FC l1 | - | - | (1, 1, 25088) | 1 | | |
| | FC l2 | - | - | (1, 1, 4096) | 1 | | |
| | FC l3 | - | - | (1, 1, 4096) | 1 | | |
| Total layer quantity (excluding Max-pooling layers) | | | | | 16 | | |
| DCNN architecture | ResNet | DCNN type | Residual network | | | | |
| Block name | Layer name (x for multiple layers) | Kernel shape | Stride shape | Feature shape | Layer quantity | | |
| | | | | | ResNet50 | ResNet101 | ResNet152 |
| | Input image | - | - | (224, 224, 3) | - | - | - |
| Conv b1 | Conv l1 | (7, 7, 64) | (2, 2) | (112, 112, 64) | 1 | 1 | 1 |
| | Max-pooling | (3, 3) | (2, 2) | (56, 56, 64) | 1 | 1 | 1 |
| Conv b2 | Conv l2-x-1 | (1, 1, 64) | (1, 1) | (56, 56, 64) | 3 | 3 | 3 |
| | Conv l2-x-2 | (3, 3, 64) | (1, 1) | (56, 56, 64) | 3 | 3 | 3 |
| | Conv l2-x-3 | (1, 1, 256) | (1, 1) | (56, 56, 256) | 3 | 3 | 3 |
| Conv b3 | Conv l3-x-1 | (1, 1, 128) | first (2, 2), rest are (1, 1) | (28, 28, 128) | 4 | 4 | 8 |
| | Conv l3-x-2 | (3, 3, 128) | (1, 1) | (28, 28, 128) | 4 | 4 | 8 |
| | Conv l3-x-3 | (1, 1, 512) | (1, 1) | (28, 28, 512) | 4 | 4 | 8 |
| Conv b4 | Conv l4-x-1 | (1, 1, 256) | first (2, 2), rest are (1, 1) | (14, 14, 256) | 6 | 23 | 36 |
| | Conv l4-x-2 | (3, 3, 256) | (1, 1) | (14, 14, 256) | 6 | 23 | 36 |
| | Conv l4-x-3 | (1, 1, 1024) | (1, 1) | (14, 14, 1024) | 6 | 23 | 36 |
| Conv b5 | Conv l5-x-1 | (1, 1, 512) | first (2, 2), rest are (1, 1) | (7, 7, 512) | 3 | 3 | 3 |
| | Conv l5-x-2 | (3, 3, 512) | (1, 1) | (7, 7, 512) | 3 | 3 | 3 |
| | Conv l5-x-3 | (1, 1, 2048) | (1, 1) | (7, 7, 2048) | 3 | 3 | 3 |
| FC block | Avg-pooling | (7, 7) | - | (1, 1, 2048) | 1 | 1 | 1 |
| | FC layer | - | - | (1, 1, 1000) | 1 | 1 | 1 |
| Total layer quantity (excluding Avg/Max-pooling layers) | | | | | 50 | 101 | 152 |

**Table 3.** Summary of model performance in predicting mechanical properties (ultimate tensile strength, UTS, and elongation to fracture, EF) for different image datasets and using different DCNN architectures. P-values are presented as 0 if they are less than 0.0001.

| Input data type | DCNN architecture | Target property | Mean absolute error, MAE | | | Average prediction accuracy (%) | Significance analysis | | |
|---|---|---|---|---|---|---|---|---|---|
| | | | unit | mean | variation | | r2-value | t-value | p-value |
| Dataset of image without contour | VGG16 | UTS | MPa | 18.42 | 14.87 | 98.2 | 0.78 | 16.8 | 0 |
| | ResNet50 | UTS | MPa | 16.81 | 13.59 | 98.3 | 0.83 | 19.7 | 0 |
| | ResNet101 | UTS | MPa | 16.18 | 11.61 | 98.4 | 0.86 | 22.0 | 0 |
| | ResNet152 | UTS | MPa | 17.45 | 12.46 | 98.3 | 0.84 | 20.4 | 0 |
| | VGG16 | EF | % | 1.12 | 1.15 | 87.0 | 0.88 | 28.5 | 0 |
| | ResNet50 | EF | % | 0.85 | 0.82 | 90.0 | 0.93 | 38.5 | 0 |
| | ResNet101 | EF | % | 0.79 | 0.67 | 90.6 | 0.95 | 45.1 | 0 |
| | ResNet152 | EF | % | 1.13 | 1.07 | 88.8 | 0.92 | 35.7 | 0 |
| Dataset of image with contour | VGG16 | UTS | MPa | 15.23 | 11.71 | 98.5 | 0.85 | 21.0 | 0 |
| | ResNet50 | UTS | MPa | 13.62 | 10.35 | 98.7 | 0.88 | 24.1 | 0 |
| | ResNet101 | UTS | MPa | 12.98 | 10.05 | 98.7 | 0.89 | 25.2 | 0 |
| | ResNet152 | UTS | MPa | 13.44 | 10.65 | 98.7 | 0.87 | 23.2 | 0 |
| | VGG16 | EF | % | 0.75 | 0.63 | 90.9 | 0.95 | 48.7 | 0 |
| | ResNet50 | EF | % | 0.69 | 0.67 | 92.4 | 0.95 | 46.8 | 0 |
| | ResNet101 | EF | % | 0.61 | 0.58 | 93.1 | 0.96 | 54.0 | 0 |
| | ResNet152 | EF | % | 0.69 | 0.67 | 92.4 | 0.95 | 46.8 | 0 |

**Table 4.** Summary of model performance in predicting mechanical properties for different baseline feature extraction method (transfer versus non-transfer image processing) and regression model (linear regression, LR, support vector machine, SVM, and random forest, RF) combinations. All the models were tested using the full image dataset (with contours). P-values are presented as 0 if they are less than 0.0001.

| Input data type | Feature extraction architecture | Target property | Model | Mean absolute error, MAE | | | Average prediction accuracy (%) | Significance analysis | | |
|---|---|---|---|---|---|---|---|---|---|---|
| | | | | unit | mean | variation | | r2-value | t-value | p-value |
| Dataset of image with contour | Transfer, ResNet101 | \multicolumn{8}{c}{Optimal NN-based regression model performances} | | | |
| | | UTS | NN | MPa | 12.98 | 10.05 | 98.7 | 0.89 | 25.2 | 0 |
| | | EF | | % | 0.61 | 0.58 | 93.1 | 0.96 | 54.0 | 0 |
| | | \multicolumn{8}{c}{Baseline regression model performances} | | | |
| | | UTS | LR | MPa | 32.04 | 8.75 | 96.9 | 0.49 | 8.8 | 0 |
| | | EF | | % | 1.44 | 0.86 | 75.5 | 0.85 | 25.1 | 0 |
| | | UTS | SVM | MPa | 20.35 | 19.10 | 98.0 | 0.69 | 13.4 | 0 |
| | | EF | | % | 1.26 | 1.36 | 71.4 | 0.82 | 22.7 | 0 |
| | | UTS | RF | MPa | 19.47 | 10.13 | 98.1 | 0.80 | 18.0 | 0 |
| | | EF | | % | 1.11 | 0.86 | 79.9 | 0.91 | 33.4 | 0 |
| | Non-transfer, PCA | UTS | LR | MPa | 37.68 | 26.50 | 96.3 | 0.01 | 0.8 | 0.42 |
| | | EF | | % | 3.31 | 2.46 | 20.1 | 0.52 | 11.0 | 0 |

**Acknowledgments**

The authors are grateful for Qihang Qiu's and Zhaotong Lu's assistance in ML model training, testing, and hyperparameter optimization.

**Data availability statement**

All relevant data are available from the authors; see details in repository posted in Zenodo (zenodo.org/record/7872513).

**Declaration of Competing Interest**

The authors declare no competing financial interests or personal relationships for the completion of work and publication of this paper.

**Funding information**

The financial support from a Pennsylvania State University College of Engineering Multidisciplinary Seed Grant is gratefully acknowledged. J.D.S. acknowledges support from the Department of Energy National Nuclear Security Administration Stewardship Science Graduate Fellowship, provided under cooperative agreement number DENA0003960.

**Supplementary material for: An image-based transfer learning approach for using *in situ* processing data to predict laser powder bed fusion additively manufactured Ti-6Al-4V mechanical properties**

Qixiang Luo[1], John D. Shimanek[1], Timothy W. Simpson[2,3], and Allison M. Beese[1,3,*]

[1] Department of Materials Science and Engineering, Pennsylvania State University, University Park, PA 16802

[2] Department of Industrial and Manufacturing Engineering, Pennsylvania State University, University Park, PA 16802

[3] Department of Mechanical Engineering, Pennsylvania State University, University Park, PA 16802

[*] Corresponding author, email: amb961@psu.edu

## 1 Sample Fabrication Details

An outer contour, with a laser power of 275 W and scan speed of 800 mm/s, was used for each sample to minimize the impact of surface roughness variations among samples and to limit differences between samples to variations within the interior bulk material. The interior region of each sample was fabricated using parallel laser passes with a hatch rotation of 245° between subsequent layers. The interior of each sample was fabricated with a single combination of power and velocity as given in tabular form **(Tab. 1)**. After fabrication, a stress-relief of 5 hours at 670 °C was applied, followed by furnace cooling for 12 h in an Argon environment. No additional heat treatment or thermal post-processing was applied to the samples prior to testing.

## 2 Mechanical Testing Details

Samples were tested under uniaxial tension at room temperature, using an electromechanical load frame (MTS Criterion Model 45) with a 150kN load cell, at a quasistatic strain rate of $3.7 \times 10^{-4}$/s. The gauge region was painted with black speckles on a white basecoat, and digital image correlation (DIC) was used to capture the evolving surface deformation fields. A 24mm vertical virtual extensometer was used for all axial strain measurements.

The distribution of the measured mechanical properties (ultimate tensile strength and elongation to fracture) as a result of the processing variations are shown in **Figure 1 (Fig. 1b, 1c)**. The processing pores were primary contributors to the resultant mechanical properties of the specimens where both very high and very low laser energy density input results in keyholing and lack-of-fusion pores, respectively, while beading up results in pores in the high Pv regime [1]. To visualize local intensity variations that indicate processing defects, the raw photodiode signal data were processed to create 2-dimensional (2D) image datasets, as shown in **Figure 4**. The time-dependent laser point data **(Fig. 4a)** were transformed into a bitmap with resolution of 10 μm per pixel, 800 x 800 pixels, where the intensity of each pixel was taken to be the average of all data points located inside that pixel. This provided a consistent signal density for all samples, regardless of laser scan speed. To correlate the properties of samples to the *in situ* processing data, 500 layers in the sample gauge region were studied as the volume of interest **(Fig. 2)**. The layer-wise signal data were separated into individual datasets for each sample, with the photodiode signals for each layer of each sample visualized into a 2D image using an 8-bit 3-channel RGB color map. This provided a total of 60,000 images for the whole dataset.

Consider the distribution of the measured mechanical properties for each specimen **(Fig. 1b, 1c)**, both distributions show a concentration of observations around their center, the minimum physical value for elongation to failure is much closer to the center of its distribution than is the case for ultimate tensile strength. Therefore, while all elongations to fracture fall within 0–17%, four of the measured ultimate tensile strengths fall below 900 MPa, with the rest of the 116 strength measurements bounded within 950–1100 MPa. To avoid overfitting biases, where data-based models favor performance within ranges containing the most measurements [2], samples with an ultimate tensile strength lower than 900MPa were excluded from the dataset.

# 3 Modeling Details

## 3.1 Data preprocessing

Before feature extraction, the images were first compressed into a default image size of 224 pixels × 224 pixels × 3 colors (RGB) for consistency with the selected DCNN architectures. To ensure a robust training and testing approach, the entire dataset was shuffled and split into five subsets, with four of those subsets used for training (80% of the data) and one for testing (20% of the data). Five-fold cross-validation was used to ensure an average model performance evaluation with minimal uncertainty caused by biased data sampling. Five total iterations were taken for each cross-validation test.

## 3.2 Transfer learning

Feature extraction was performed by the processing module of an image-based ML workflow, the details of which are given in **Table S1**. Spatial photodiode signal intensity **(Fig. 4c, 4d)** was condensed by the DCNN architectures from 2D pixel matrices into 1D feature vectors. Each feature vector represents the information contained within the original image, which relates to the in-process heat distribution and defect formation, without the expense and inherent variability of manually labeling images and training a CNN from scratch. The multiple convolutional blocks of the pre-tuned DCNN architectures, the details of which are given in **Table 2**, allowed for the extraction of image features like hotspots [3,4], although the physical interpretability of each feature vector is not necessary for their use in the mechanical property regression models of the present study.

This information extraction process was enabled by the transferring of pre-trained DCNN architectures trained to achieve success with the large-scale image database ImageNet [5], which has 1000 categories and over 1.2M images of normal life objects. Example images from ImageNet training are shown in **Figure 4**, and, by extracting and summarizing the general image characteristics (edges and hotspots) through continuous CNN blocks, vectorized fingerprints of each image are made available to simple classification or regression models. The practice of transfer learning takes the tuned DCNN models, as trained on large-scale image recognition tasks, and applies their feature extraction capabilities to similar image-based problems in other domains. Although the classification tasks in the computer vision field are very different from the requirements of other domains, transfer learning in this case refers only to the feature extraction process and not to the task-specific downstream model. In the present context, these pre-trained DCNN image encoder models were therefore used to process the reconstructed signal images into information-dense feature vectors with length 2048 or 4096, depending on the DCNN architecture. The application of transfer learning thereby allowed for the computationally efficient generation of representative signal features for each processing condition considered. Moreover, the adoption of a widely used computer vision protocol advances standardization in the implementation of ML tools for *in situ* AM signal processing, whose speed and flexibility make them promising components for real-time process diagnosis and closed-loop correction methods.

3.3 Model performance

As shown in shown in **Figure 8**, comparing the DCNN structures, ResNet architecture design has reduced computational complexity compared to VGG design, where the deeper ResNet architecture (ResNet50, 50 layers) ran faster (8 ms/image) than the shallow VGG architecture (VGG16, 16 layers, 10 ms/image) using the GPU resources described in **Table S1**. This agreed with literature investigations on the model complexity, quantified via number of floating-point operations (FLOPs), and computation time, via inference time per image input, by reporting a computational complexity reduced by a factor of three and 35% fewer computations on ResNet50 compared to VGG16 [6,7].

Additionally, the proposed neural network (NN) regression model showed promising performance metrics for future application to real-time process control. Despite the greater complexity of the neural network model relative to simpler regression models tested as baseline comparisons, it is important to note that its accuracy does not necessarily come at the cost of prohibitive computational time since it is well supported by GPU hardware acceleration. For example, the average training times per image were 0.1 ms for LR, 21 ms for SVM, and 40 ms for RF using central processing unit (CPU) computation. The NN was comparably slow, at 320 ms on CPU, but can be accelerated to 8 ms on GPU without loss of accuracy.

As mentioned in the main text, the learning curves of the NN regression model are shown in **Figure S1** and show no signs of underfitting or overfitting.

**Supplementary Figures**

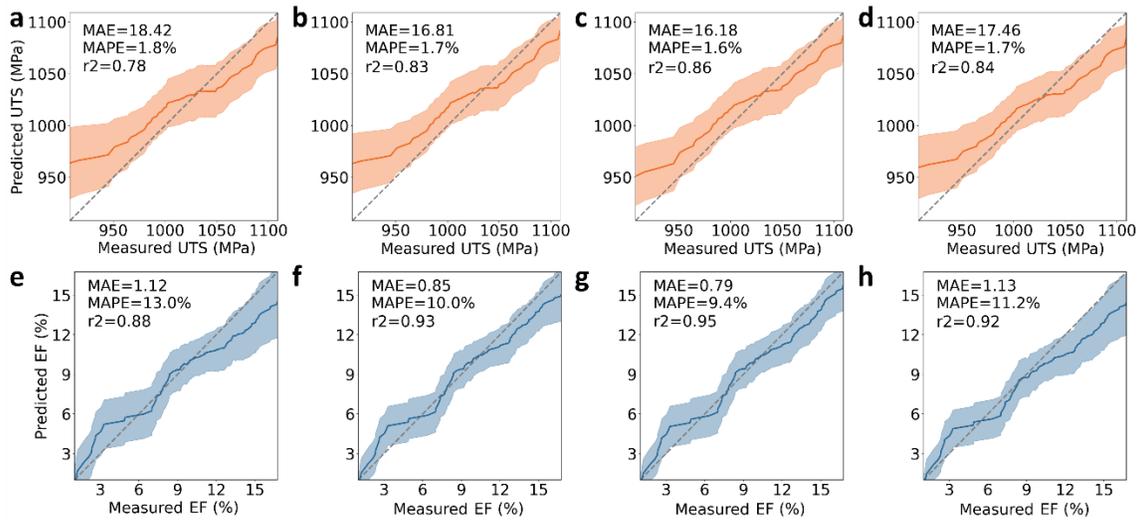

**Figure S1.** Performance of regression models in terms of predicting the measured ultimate tensile strength and elongation to failure based on the *in situ* processing image dataset without contours. (a-d) Predictions of ultimate tensile strength and (e-h) predictions of elongation to fracture using different DCNN architectures: (a, e) VGG16, (b, f) ResNet50, (c, g) ResNet101, and (d, h) ResNet152. The solid lines and the shaded bands correspond to the predicted average mechanical properties of each sample and the prediction variations for each single build layer within the sample, respectively. The closer the prediction results are to the gray dashed lines representing perfect agreement between measured results and predicted properties, the better the prediction.

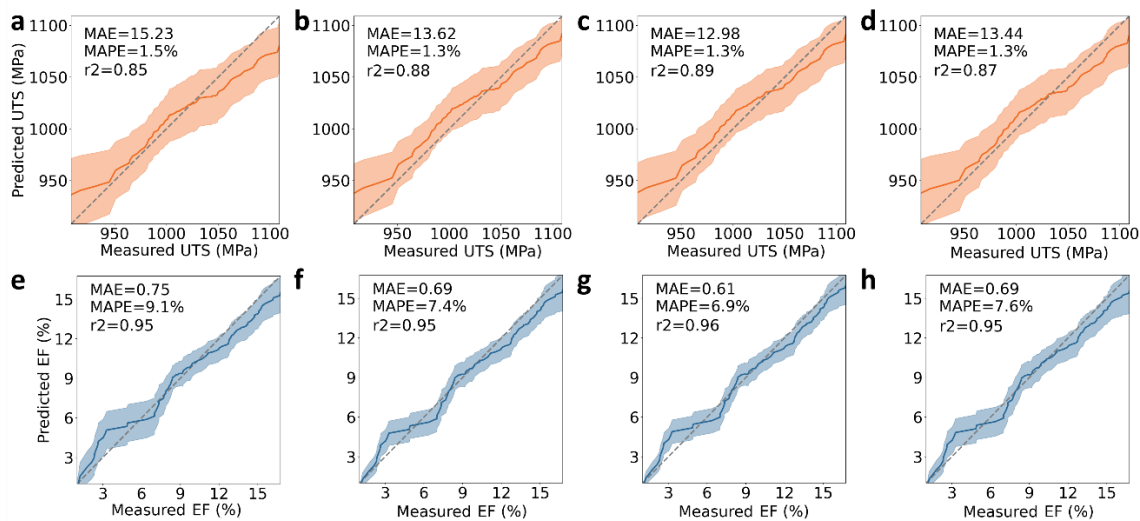

**Figure S2.** Performance of regression models in terms of predicting the measured ultimate tensile strength and elongation to failure based on the *in situ* processing image dataset with contours. (a-d) Predictions of ultimate tensile strength and (e-h) predictions of elongation to fracture using different DCNN architectures: (a, e) VGG16, (b, f) ResNet50, (c, g) ResNet101, and (d, h) ResNet152.

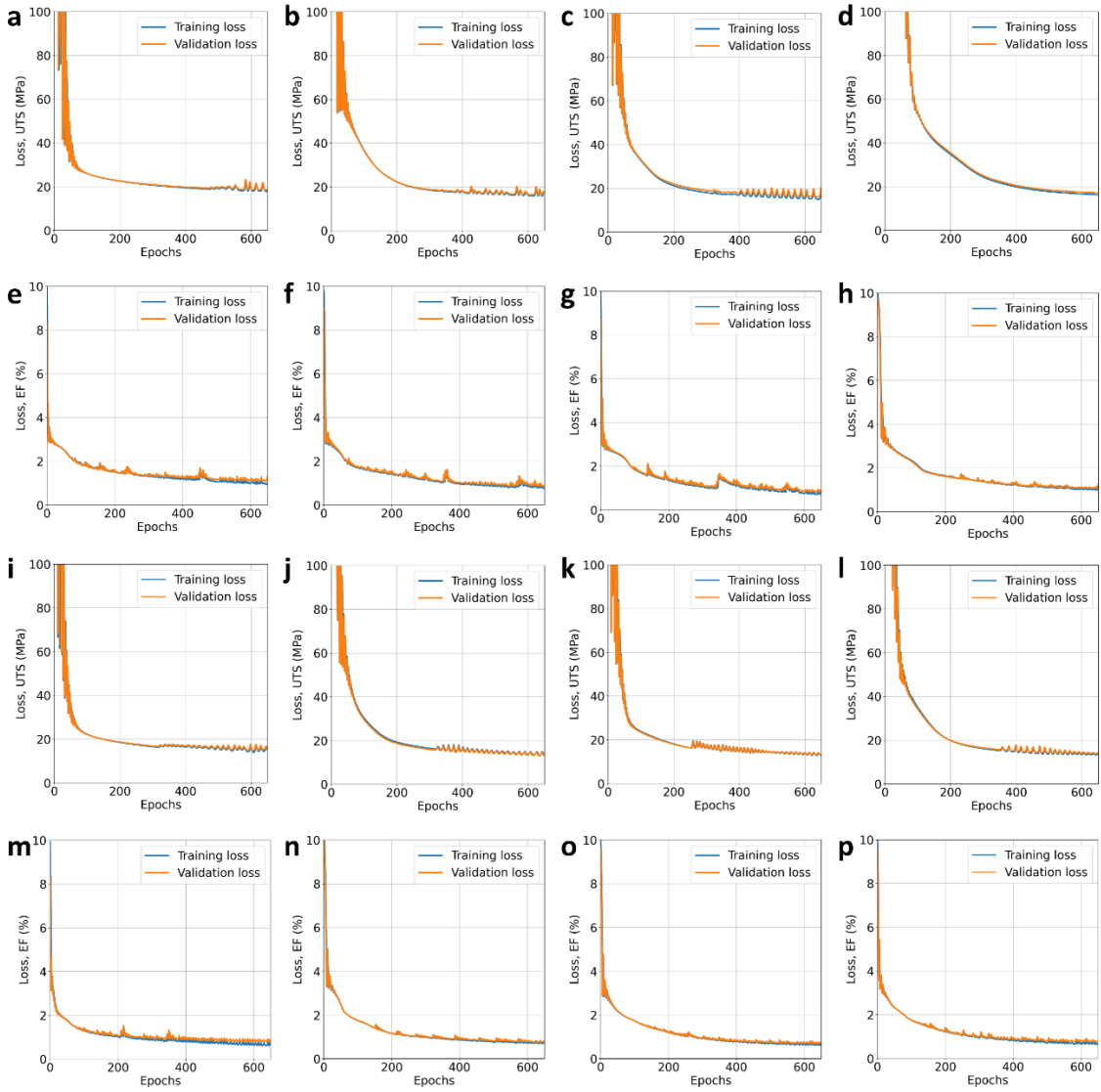

**Figure S3.** Learning curves for all test cases using the no contour (a-h) and contour (i-p) datasets with VGG16 (a, e, I, m), ResNet50 (b, f, j, n), Resnet101 (c, g, k, o), and ResNet152 (d, h, l, p) architectures for predicting UTS (a-d, i-l) and EF (e-h, m-p).

# Supplementary Tables

**Table S5**. Summary of computational and model details used in the presents study, including computational resources (device configuration and programming/software information), DCNNs (architecture), and self-designed regression model (architecture and hyperparameters).

| Computational resources | | | | | |
|---|---|---|---|---|---|
| Device 1: local desktop | | | Device 2: Google Colab | | |
| GPU | NVIDIA GeForce RTX 2070, 8GB | | GPU | NVIDIA Tesla T4 or P100 16GB | |
| CPU | Intel Core i7-8700K, 6 cores | | CPU | 2 × Intel Xeon CPU @2.20 GHz | |
| RAM | 64GB | | RAM | 52GB | |
| Programming packages and software | | | | | |
| Python | v3.8.8 | TensorFlow | v1.12.0 | Keras | v2.2.4 |
| scikit-learn | v0.24.1 | Numpy | v1.20.1 | Matplotlib | v3.3.4 |
| CUDA Toolkit | v10.0 | cuDNN | v7.3.1 | | |
| Modelling part I: feature extraction and DCNNs | | | | | |
| Architecture | Network structure | Input shape (pixel) | Output layer, extracted feature length (pixel) | Depth | Number of parameters |
| VGG16 | Plain | 224 x 224 x3 | fc2, 4096 | 16 | 138.4M |
| ResNet50 | Residual | 224 x 224 x3 | avg_pool, 2048 | 50 | 25.6M |
| ResNet101 | Residual | 224 x 224 x3 | avg_pool, 2048 | 101 | 44.7M |
| ResNet152 | Residual | 224 x 224 x3 | avg_pool, 2048 | 152 | 60.4M |
| Modelling part II: regression model | | | | | |
| Sequential model architecture | | | | Model hyperparameter | |
| Layer | Input shape (pixel) | Output shape (pixel) | Activation function | Optimizer | Adam |
| Dense (fc1) | 4096 or 2048 | 1024 | ReLu | Learning rate | 0.001 |
| Dense (fc2) | 1024 | 512 | ReLu | Loss function | MAPE |
| Dense (fc3) | 512 | 256 | ReLu | Epoch | 650 |
| Dense (fc4) | 256 | 64 | ReLu | Others | Default (Keras) |
| Dense (fc5) | 64 | 16 | ReLu | Cross-validation parameter | |
| Dense (fc6) | 16 | 8 | ReLu | Train-test set | 80-20%, 5-fold |
| Dense (fc7) | 8 | 1 | Linear | Shuffle | True |
| Summary | Number of parameters | 4.9M (VGG16) | 2.8M (ResNet) | | |